\documentclass[superscriptaddress,aps,prb,preprint]{revtex4-2}
\usepackage{bm}
\usepackage{dcolumn}
\usepackage{graphicx}
\usepackage{hyperref}
\usepackage{mathrsfs}
\usepackage{multirow}
\usepackage{rotating}
\usepackage{url}
\usepackage{color}
\usepackage{amsmath}
\usepackage{amssymb}
\newcommand{\etal}{{\em et al}.\ }

\begin{document}


\title{Switchable band topology and geometric current in sliding bilayer elemental ferroelectric}

\author{Zhuang Qian}
\affiliation{Key Laboratory for Quantum Materials of Zhejiang Province, Department of Physics, School of Science and
Research Center for Industries of the Future, Hangzhou Zhejiang 310030, China}
\affiliation{Institute of Natural Sciences, Westlake Institute for Advanced Study, Hangzhou, Zhejiang 310024, China}
\author{Zhihao Gong}
\affiliation{ZJU-Hangzhou Global Scientific and Technological Innovation, School of Physics, Zhejiang University, Hangzhou 311215, China}
\author{Jian Li}
\affiliation{Key Laboratory for Quantum Materials of Zhejiang Province, Department of Physics, School of Science and
Research Center for Industries of the Future, Hangzhou Zhejiang 310030, China}
\affiliation{Institute of Natural Sciences, Westlake Institute for Advanced Study, Hangzhou, Zhejiang 310024, China}
\author{Hua Wang}
\email{daodaohw@zju.edu.cn}
\affiliation{ZJU-Hangzhou Global Scientific and Technological Innovation, School of Physics, Zhejiang University, Hangzhou 311215, China}
\author{Shi Liu}
\email{liushi@westlake.edu.cn}
\affiliation{Key Laboratory for Quantum Materials of Zhejiang Province, Department of Physics, School of Science and
Research Center for Industries of the Future, Hangzhou Zhejiang 310030, China}
\affiliation{Institute of Natural Sciences, Westlake Institute for Advanced Study, Hangzhou, Zhejiang 310024, China}

%

\date{\today}


\begin{abstract}
We demonstrate that sliding motion between two layers of the newly discovered ferroelectric and topologically trivial bismuth (Bi) monolayer [Nature 617, 67 (2023)] can induce a sequence of topological phase transitions, alternating between trivial and nontrivial states. Interestingly, a lateral shift, even when preserving spatial symmetry, can still switch the quantum spin Hall state on and off. The substantial band-gap modulation and band inversion due to interlayer sliding arise primarily from the intralayer in-plane charge transfer processes involving Bi atoms at the outermost atomic layers, rather than the interlayer charge redistribution. 
We map out the topological phase diagram and the geometric Berry curvature-dipole induced nonlinear anomalous Hall response resulting from sliding, highlighting the potential for robust mechanical control over the edge current and the Hall current. Bilayer configurations that are $\mathbb{Z}_2$ nontrivial can produce drastically different transverse currents orthogonal to the external electric field. This occurs because both the direction and magnitude of the Berry curvature dipole at the Fermi level depend sensitively on the sliding displacement.  Our results suggest that bilayer bismuth could serve as a platform to realize power-efficient ``Berry slidetronics" for topology memory applications. 
\end{abstract}

\pacs{
}
\maketitle

\newpage

Two-dimensional (2D) van der Waals (vdW) materials have garnered significant attention in the fields of condensed matter physics and materials science~\cite{Duong17p11803, Burch18p47, Wu21p9229, Wang23p542}. Their unique properties, such as highly tunable electronic properties and remarkable mechanical flexibility, position them as promising candidates for a myriad of applications, ranging from next-generation electronics and optoelectronics to energy storage and flexible devices~\cite{Wu13p081406, Feng12p866, Nahas20p5779, Jiang23p256902, Sauer23p35}. The layers in these materials are held together by weak vdW forces, making them particularly susceptible to twisting and sliding. Such structural deformations, especially twisting,  can profoundly alter the electronic behaviors of 2D materials~\cite{Yankowitz19p1059, Sharpe19p605, Liu20p221}. For example, adjusting the twist angle between adjacent layers results in the formation of long-wavelength moir\'{e} superlattices. The moir\'{e} potential with a modulation period much larger than the atomic lattice constant can quench the kinetic energy of electrons and effectively enhance the electronic correlation, setting the stage for various quantum phenomena, including Hofstadter butterfly pattern~\cite{Bistritzer11p035440, Wang12p3833}, superconductivity~\cite{Cao18p43, Chen19p215, Yankowitz19p1059}, Mott insulators~\cite{Chen19p237, Shimazaki20p472, Cao18p80}, and moir\'{e} excitons~\cite{Jin19p76, Tran19p71}.

Recent studies have demonstrated that interlayer sliding can induce out-of-plane polarization in bilayer systems, even when they are composed of nonpolar monolayers~\cite{Li17p6382, Ji23p146801, Miao22p1158}. This phenomenon is attributed to changes in the layer stacking configuration caused by sliding, which leads to a reversal in the direction of charge transfer between the layers and consequently switchable out-of-plane polarization~\cite{Li17p6382}. Notable bilayer systems possessing sliding ferroelectricity that have been confirmed experimentally are bilayer graphene and transition metal dichalcogenides~\cite{Yasuda21p6549, ViznerStern21p1462, Zheng20p71, Roge22p973}.
Unlike twisting, sliding in bilayers of identical monolayers generally has a less significant impact on their electronic and optical properties, as it doesn't significantly modulate the periodic potential landscape. One notable exception, predicted theoretically, occurs in sliding bilayer graphene, where the electronic topological transition involving pair annihilations of massless Dirac fermions is sensitively influenced by the direction of the lateral interlayer shift~\cite{Son11p155410}. 
However, sliding-induced electronic topological transitions~\cite{Ren22p235302}, such as the $\mathbb{Z}_2$ topological phase transition, are rarely reported in gapped 2D materials. This scarcity isn't particularly surprising. A 
$\mathbb{Z}_2$ topological phase transition, characterized by a change in the band structure's topology, demands the closure of the band gap, wherein the conduction and valence bands must touch at least at one point in the momentum space.  The intrinsic inert response of the band gap to sliding naturally hinders the transition between a trivial insulator and a topological insulator. In addition, topological insulators typically exhibit a larger dipole transition matrix element and Berry curvature compared to trivial insulators~\cite{xu20p6119}, making them promising candidates for Berry curvature memory~\cite{Xiao20p1028} based on the geometric Berry curvature-dipole induced nonlinear anomalous Hall effect. In particular, Berry curvature memory, deriving its functionality from the topological properties of materials, could offer advantages in terms of speed, energy efficiency, and durability over conventional semiconductor memory technologies~\cite{wang19p119, Xiao20p1028, Du20p022025, Singh20p046402}. A sliding-induced $\mathbb{Z}_2$ topological phase transition, if achievable, has the potential to serve as an additional controlling knob, unlocking novel functionalities and device types.

In this study, we utilize first-principles density functional theory (DFT) calculations including hybrid functionals to demonstrate that sliding motion between two layers of the newly discovered single-element ferroelectric bismuth (Bi) monolayer~\cite{Gou23p67} triggers successive trivial-to-nontrivial topological phase transitions, accompanied by nonlinear anomalous Hall effects. We find that, although bismuth monolayer is topologically trivial, stacking it into a bilayer and applying appropriate sliding can transform it into a quantum spin Hall insulator.
This sliding-tunable band gap and the ensuing quantum phase transition could enable energy-efficient mechanical switching between on and off states of a quantum spin Hall insulator, along with its quantized edge conductance. An intriguing aspect is that the substantial modulation of the band gap and the resulting band inversion, driven by interlayer sliding, largely originate from the intralayer in-plane charge transfer of the outermost Bi atoms. 
We have constructed a topological phase diagram as a function of in-plane sliding motions of various magnitudes along all possible directions, providing a precise map for the fine-tuning of quantum states. Additionally, we show that $\mathbb{Z}_2$ nontrivial states induced by different magnitudes of sliding motions can generate nonlinear Hall currents that vary in both direction and magnitude.  This suggests another sliding-tunable current, in addition to the symmetry-protected edge current. We propose that bilayer bismuth may serve as a promising platform for the realization of power-efficient “Berry slidtronics", harnessing both deterministically controllable band topology and geometric current for nonvolatile multistate memory applications.

All plane-wave DFT calculations are performed using \texttt{Quantum Espresso}~\cite{Giannozzi09p395502, Giannozzi17p465901}. The vdW bilayers are fully optimized using a generalized gradient approximation of the Perdew-Burke-Ernzerhof (PBE) type~\cite{Perdew96p3865}, augmented with Grimme dispersion corrections, and employing Garrity-Bennett-Rabe-Vanderbilt (GBRV) ultrasoft pseudopotentials~\cite{Garrity14p446}. We use a plane-wave kinetic energy cutoff of 50 Ry, a charge density cutoff of 250 Ry, a 12$\times$12$\times$1 Monkhorst-Pack $k$-point mesh for Brillouin zone (BZ) integration, an ionic energy convergence threshold of 10$^{-5}$ Ry, and a force convergence threshold of 10$^{-4}$ Ry to converge the structural parameters.  
For band structure calculations, the spin-orbit coupling (SOC) is taken into account at the fully relativistic level with norm-conserving pseudopotentials provided by the PseudoDoJo project~\cite{Van18p39}, in combination with energy cutoffs of 80 Ry and 320 Ry for the wavefunction and charge density, respectively. 
Additionally, we compute the electronic structure employing the Heyd-Scuseria-Ernzerhof (HSE) hybrid functional~\cite{Krukau06p224106} with a $4\times 4\times 1$ $q$-point grid during the self-consistent-field cycles. The HSE band structure is obtained through Wannier interpolation~\cite{Marzari12p1419} using \texttt{Wannier90}~\cite{Pizzi20p165902} interfaced with \texttt{Quantum Espresso}. The topological state is determined by computing the $\mathbb{Z}_2$ topological index using the Wilson loop method, based on the Wannier tight-binding Hamiltonians constructed with \texttt{Wannier90}.

\begin{figure}[h]
    \centering
    \includegraphics[width=0.8\textwidth]{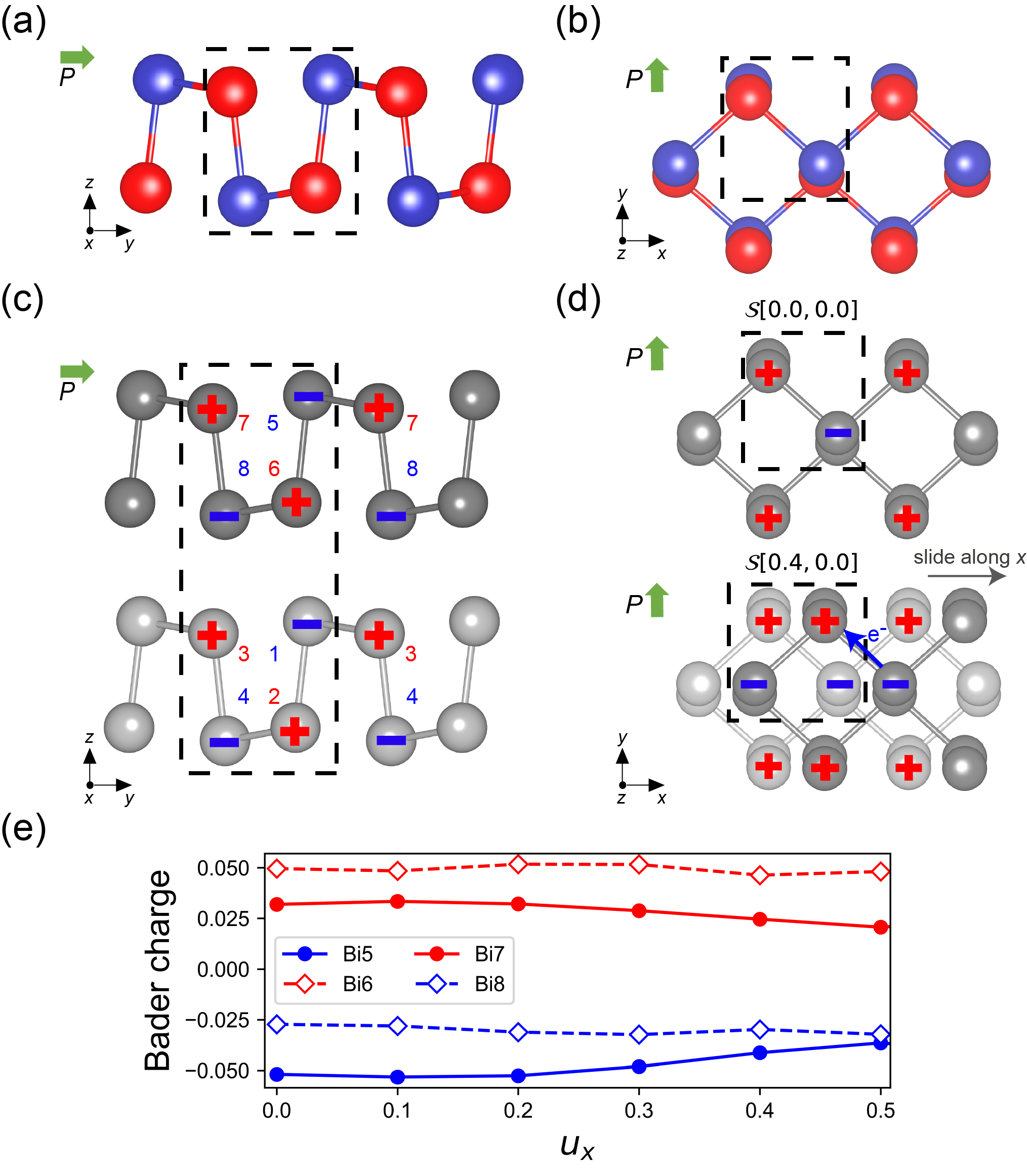}
    \caption{ Schematics of the (a) side and (b) top view of ferroelectric monolayer bismuth with in-plane polarization (green arrow) along the $y$-axis. The atoms are colored based on the Bader charge analysis (blue for negative and red for positive). The monolayer has a space group of $Pmn2_1$. (c) Schematic of bilayer bismuth that has one monolayer stacked directly atop another. The positive and negative Bader charges are labeled with $+$ and $-$ signs, respectively. The Bi atoms in the upper layer are colored in black, while those in the lower layer are in gray. (d) Top view of configuration $\mathcal{S}[0.0, 0.0]$ (top) and configuration $\mathcal{S}[0.4, 0.0]$ (bottom), with the upper layer shifted by $\vec{d}_s=0.4\vec{a}$.  The blue arrow represents the direction of electron transfer during the sliding process. (e) Bader charge evolution of Bi atoms as a function of $u_x$.}
    \label{structure}
\end{figure}

The structure of ferroelectric bismuth monolayer in the space group of $Pmn2_1$ can be viewed as a distorted, phosphorene-like structure, as depicted in Fig.~\ref{structure}(a) and (b). The presence of out-of-plane atomic buckling breaks the inversion symmetry, leading to a spontaneous charge disproportion between neighboring Bi atoms and an in-plane polarization along the $y$ axis~\cite{Xiao18p1707383}. 
Xu~\etal~have recently confirmed experimentally the switchable in-plane polarization in this single-element 2D material~\cite{Gou23p67}.
The bilayer configuration studied in this work consists of one monolayer stacked directly atop another [see Fig.~\ref{structure}(c)]. In this vertically stacked configuration that preserves the symmetry of the individual monolayers, the lattice vectors $(\vec{a}, \vec{b})$ of the upper layer are oriented parallel to those of the lower layer. Upon sliding the upper layer relative to the lower one, the sliding vector ($\vec{d}_s$) is defined as $\vec{d}_s=u_x\vec{a}+u_y\vec{b}$, with the resulting configuration labeled as $\mathcal{S}[{u_x},{u_y}]$. For instance, consider the sliding vector $\vec{d}_s=0.4\vec{a}$ shown in Fig.~\ref{structure}(d), the configuration is denoted as  $\mathcal{S}[0.4, 0.0]$.
It is important to note that sliding in this bilayer configuration does not break the out-of-plane inversion symmetry, thereby preventing the emergence of sliding ferroelectricty featuring interlayer charge transfer-induced out-of-plane polarization. Interestingly, we observe a substantial sliding-induced intralayer charge transfer.  
By tracking the changes in the Bader charges of all Bi atoms during the lateral shift along $x$, as shown in Fig.~\ref{structure}(e), we find that the electron transfer is particularly pronounced for Bi atoms located in the outermost layers: electrons move from negatively charged Bi atoms (labeled as Bi5) to adjacent positively charged ones (labeled as Bi7). The intralayer charge transfer is accompanied by a substantial modulation of the band structure, leading to a topological phase transition.

\begin{figure}[h]
    \centering
    \includegraphics[width=0.8\textwidth]{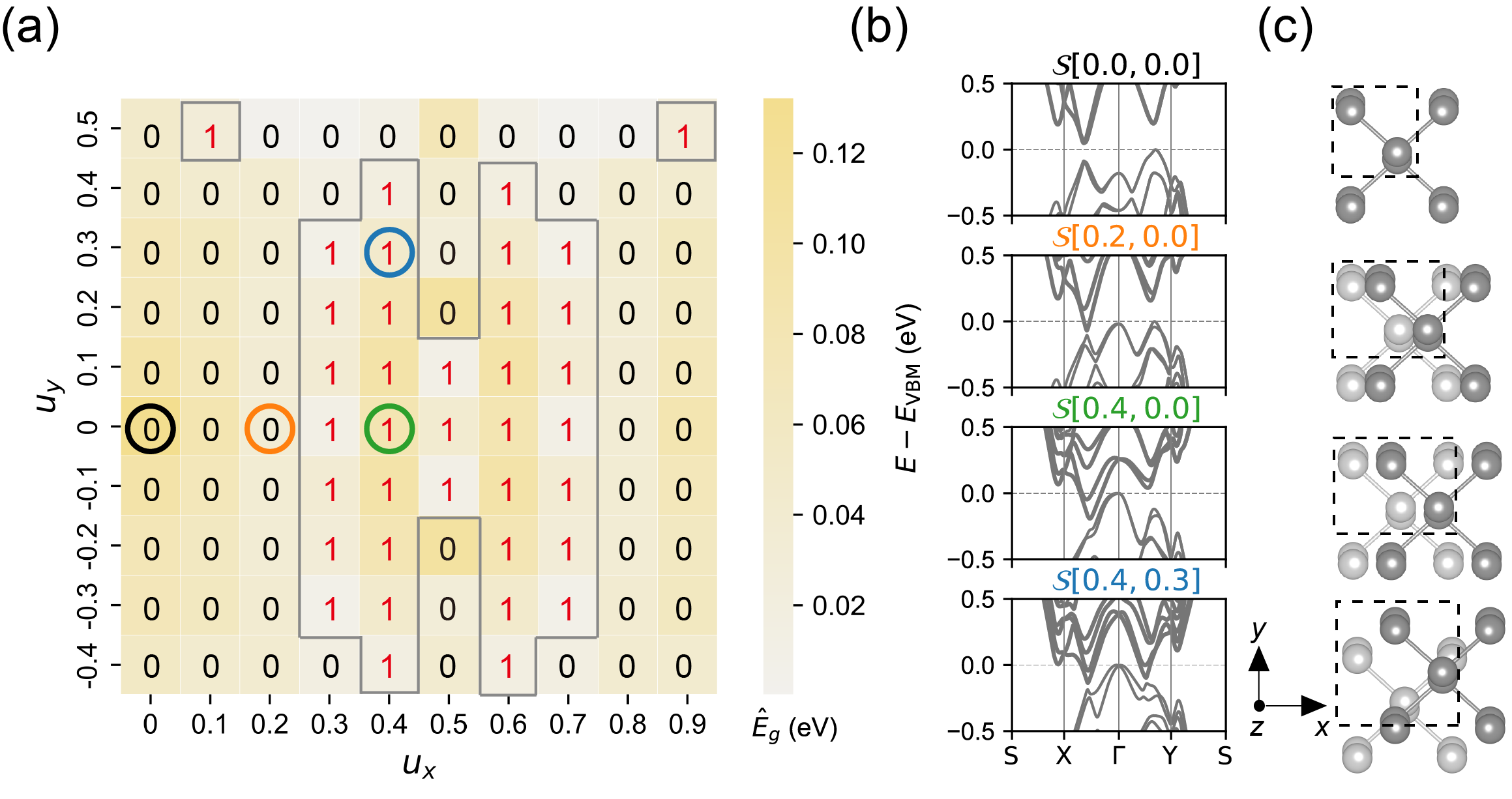}
    \caption{(a) Heatmap of topological states of bilayer  bismuth in configuration $\mathcal{S}[u_x, u_y]$, with ``0" and ``1" representing $\mathbb{Z}_2$ trivial and nontrivial band topology, respectively.  The background color scales with the value of direct band gap ($\hat{E}_g$). The gray line traces the phase boundary. (b) Band structures of configurations $\mathcal{S}[{0.0},{0.0}]$, $\mathcal{S}[{0.2},{0.0}]$, $\mathcal{S}[{0.4},{0.0}]$, and $\mathcal{S}[{0.4},{0.3}]$, highlighted in (a), with their top views are shown in (c). The valence band maximum (VBM) is chosen as the reference energy when plotting the band structure.}
    \label{heatmap}
\end{figure}

Based on our benchmark calculations comparing HSE+SOC and PBE+SOC, we find that both methods predict similar band structures (see details in Supplementary Material). Therefore, we map out the topological state of bilayer Bi  as a function of sliding defined by $u_x$ and $u_y$. 
In Fig.~\ref{heatmap}(a), a heatmap is designed to concurrently illustrate the topological state and the magnitude of the direct bandgap ($\hat{E}_g$): a ``0'' denotes a trivial state, and a ``1'' signifies a nontrivial state; the intensity of the background color is proportional to the value of $\hat{E}_g$. We note that the presence of a mirror symmetry $\mathcal{M}_x$ in monolayer renders configuration $\mathcal{S}[u_x, u_y]$ equivalent to $\mathcal{S}[1-u_x, u_y]$.
The heatmap unveils a complex landscape of topological phase transitions induced by in-plane sliding. 
A cluster of configurations displaying nontrivial band topology coalesces into a distinctive butterfly-shaped region, with its center at  $\mathcal{S}[0.5, 0.0]$. Notably, an outlier state at $\mathcal{S}[0.1, 0.5]$ is nontrivial, anomalously nestled within a cluster of trivial states. 

The heatmap provides a convenient visual tool to design sliding pathway to fine tune quantum states. Figure~\ref{heatmap}(b) presents the band structures of four representative configurations shown in Fig.~\ref{heatmap}(c). The starting configuration of bilayer Bi, $\mathcal{S}[0.0, 0.0]$ is a semiconductor with a direct band gap of approximately 0.132 (0.174) eV, as predicted by PBE+SOC (HSE+SOC). At a sliding distance of $u_x=0.2$, the gap is almost closed. Subsequent sliding along $x$ to $u_x=0.4$ reopens the band gap. It is confirmed with both HSE+SOC and PBE+SOC that before the gap closure, bilayer Bi configurations like $\mathcal{S}[0,0]$ and $\mathcal{S}[0.2,0]$ are trivial insulators with $\mathbb{Z}_2=0$, whereas $\mathcal{S}[0.4,0]$ is a topological insulator featuring a direct band gap $\hat{E}_g$ of 0.083 eV and $\mathbb{Z}_2=1$. It is clear that sliding along the $x$-axis for one lattice vector induces a sequential trivial-nontrivial-trivial topological phase transitions.  

The in-plane polarization, which breaks the inversion symmetry in bilayer ferroelectric Bi, offers a distinctive potential to harness substantial Berry curvatures associated with nontrivial states for the generation of a giant nonlinear anomalous Hall current~\cite{Jin21p9468}. Unlike the conventional Hall effect, which requires the breaking of time-reversal symmetry and scales linearly with the applied electric field \(\mathcal{E}\), the nonlinear Hall current manifests as a second-order response to \(\mathcal{E}\). In the nonlinear anomalous Hall effect, the induced geometric current in static limit flows perpendicular to the electric field and exhibits a quadratic dependence on field strength, as expressed by the equation
\begin{equation}
    j^{0}_{a} = \chi_{abc}\mathcal{E}_b\mathcal{E}_c^*.
\end{equation}
Here, the nonlinear conductivity tensor $\chi_{abc}$ is given by the equation~\cite{Sodemann15p216806, Morimoto16p245121}
\begin{equation}
    \chi_{abc} = -\epsilon_{adc}\frac{e^3\tau}{2(1+i\omega\tau)}D_{bd},
\end{equation}
where \(\epsilon_{adc}\) is the Levi-Civita tensor, \(\tau\) is the relaxation time, and \(\omega\) is the frequency of the applied external field, with \(\mathcal{E}_c(\omega) = \mathcal{E}_c e^{i\omega t}\). The nonlinear conductivity plays a crucial role and is intimately connected to the Berry curvature dipole \(D\), with the primary contribution originating from states near the Fermi surface. The specific form of Berry curvature dipole \(D_{bd}\) is expressed as
\begin{equation}
    D_{bd} = \sum_n \int_{k} f_n^0(\partial_b \Omega_n^d) = -\sum_n\int_{k} (\partial_b f_n^0) \Omega_n^d,
\end{equation}
where \(\partial_b = \frac{\partial}{\partial k_b}\), \(f_n^0\) is the equilibrium Fermi-Dirac distribution for the $n$th band, and \(\Omega_n^d\) represents the component of Berry curvature for the $n$th band along \(k_d\) in momentum space. It is a well-established principle that the simultaneous presence of time-reversal and inversion symmetries imposes a restriction of zero Berry curvature across the entire Brillouin zone, precluding the occurrence of the nonlinear Hall effect. Here, in bilayer ferroelectric Bi, the manifestation of in-plane polarization inherently gives rise to non-zero Berry curvatures, which is a prerequisite for the emergence of a nonlinear anomalous Hall current.



\begin{figure}[h]
    \centering
    \includegraphics[width=0.8\textwidth]{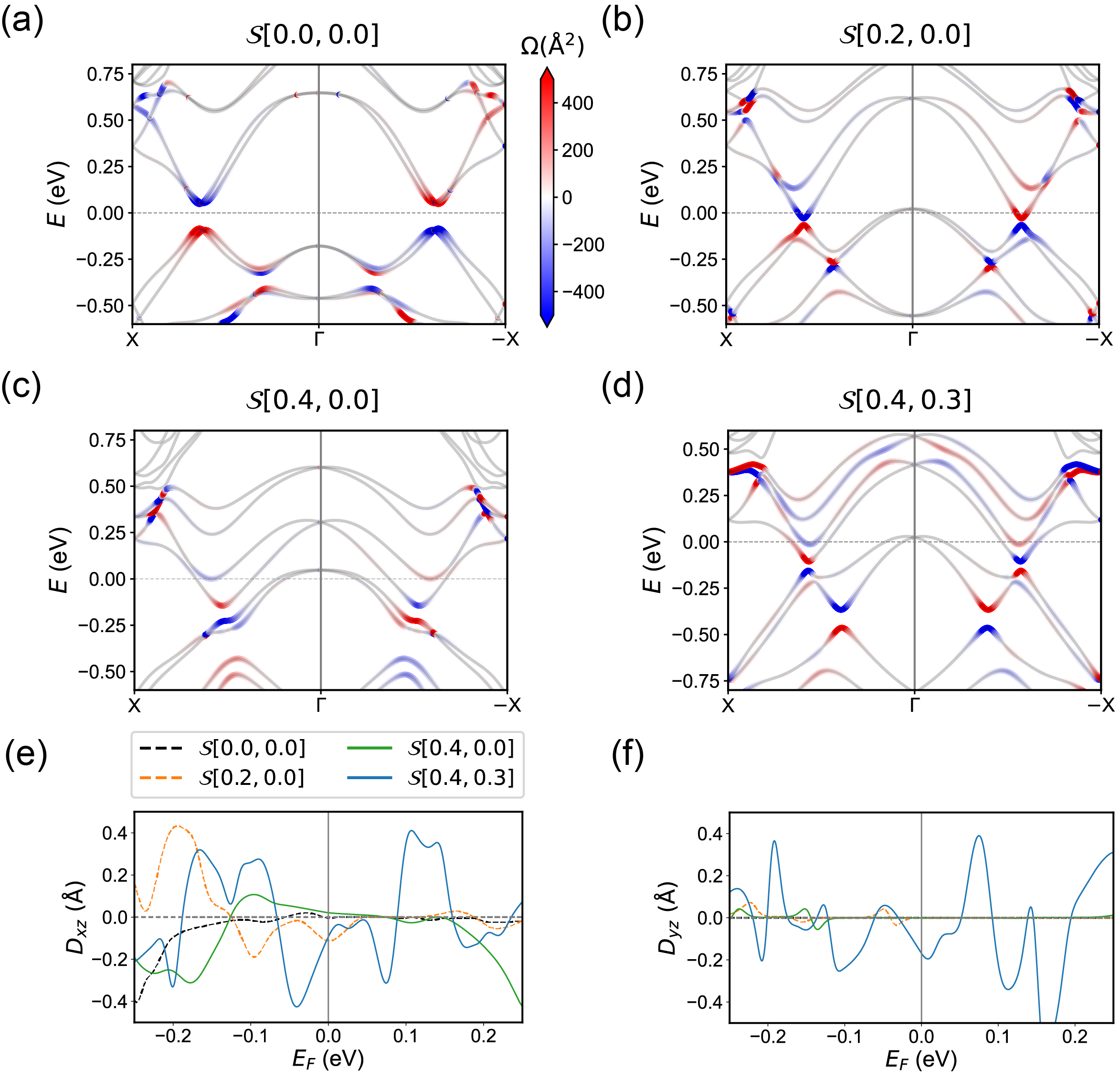}
    \caption{Berry curvature-resolved band structures for (a) $\mathcal{S}[0.0, 0.0]$, (b) $\mathcal{S}[0.2, 0.0]$, (c) $\mathcal{S}[0.4, 0.0]$, and (d) $\mathcal{S}[0.4, 0.3]$. Berry curvature dipole component of (e) $D_{xz}$ and (f) $D_{xz}$ as a function of $E_F$. For each configuration, $E_F=0$ corresponds to its own intrinsic undoped value calculated with DFT.
    }
    \label{berrycurvature}
\end{figure}

We present the Berry curvature-resolved band structures in Fig.~\ref{berrycurvature} for two topologically trivial configurations, $\mathcal{S}[0.0,0.0]$ and $\mathcal{S}[0.2,0.0]$, and two nontrivial configurations, $\mathcal{S}[0.4,0.0]$ and $\mathcal{S}[0.4,0.3]$.
These configurations exhibit distinct distributions of Berry curvatures. By comparing Fig.~\ref{berrycurvature}(a)-(b) with Fig.~\ref{berrycurvature}(c)-(d), a pronounced reversal in the sign of the Berry curvature becomes evident 
near the $k$-points of gap closure. Moreover, significant variations in the Berry curvature distribution among nontrivial configurations, such as $\mathcal{S}[0.4,0.0]$ and $\mathcal{S}[0.4,0.3]$, are also observed, as indicated in Fig.~\ref{berrycurvature}(c)-(d). The main structural difference between $\mathcal{S}[0.4,0.0]$ and $\mathcal{S}[0.4,0.3]$ is the additional sliding along the $y$-axis ($u_y=0.4$) introduced to the latter configuration. These results highlight that interlayer sliding can not only induce a topological phase transition but also greatly modulate the band structure and thus the Berry curvature distribution within the same $\mathbb{Z}_2$ state.

We further compute components of Berry curvature dipole, $D_{xz}$ and $D_{yz}$, for these representative configurations as a function of the Fermi level ($E_F$), with results presented in  Fig.~\ref{berrycurvature}(e) and (f). The nonlinear Hall current along $y$ and $x$ scales with $D_{xz}$ and $D_{yz}$, respectively. It is noted that for each configuration, $E_F=0$ corresponds to the intrinsic undoped value computed with DFT. 
The $D_{xz}$ spectra of these bilayer configurations are markedly different.
Within a relatively narrow energy window of $\pm$0.05 eV around $E_F$, the topologically trivial configuration $\mathcal{S}[0.0,0.0]$ exhibits minimal changes, while $\mathcal{S}[0.2,0.0]$ undergoes more pronounced variations. This can be understood by comparing their band structures [see Fig.~\ref{berrycurvature}(a) and (b)], where the valance bands at the $\Gamma$ point in $\mathcal{S}[0.2,0.0]$ have higher energies, contributing more states near the Fermi level. Interestingly, the nontrivial configuration of $\mathcal{S}[0.4,0.0]$ acquires positive $D_{xz}$ values within the same energy window, but further sliding along $y$ results in negative and considerably greater $D_{xz}$ values in $\mathcal{S}[0.4,0.3]$.
The value of $D_{xz}$ can reach $\approx$0.4 $\rm\AA$, larger than that of $\approx$0.1 $\rm\AA$ in WTe$_2$ and MoTe$_2$~\cite{You18p121109}.
Overall, $D_{xz}$ profiles of all configurations show strong variations in both sign and magnitude. 
This suggests that the seemingly ``gentel'' lateral shift in real space actually can cause unexpectedly large changes in the band dispersion and band inversion in momentum space. The $D_{yz}$ value of configuration $\mathcal{S}[0.0,0.0]$ is strictly zero due to the mirror symmetry $\mathcal{M}_x$. Although sliding along $x$ breaks this mirror symmetry, the values of $D_{yz}$ in $\mathcal{S}[0.2,0.0]$ and $\mathcal{S}[0.4,0.0]$ remain small. In contrast, sliding along the $y$-axis leads to giant $D_{yz}$ values in $\mathcal{S}[0.4,0.3]$, which oscillate strongly with respect to $E_F$.

\begin{figure}[h]
    \centering
    \includegraphics[width=0.8\textwidth]{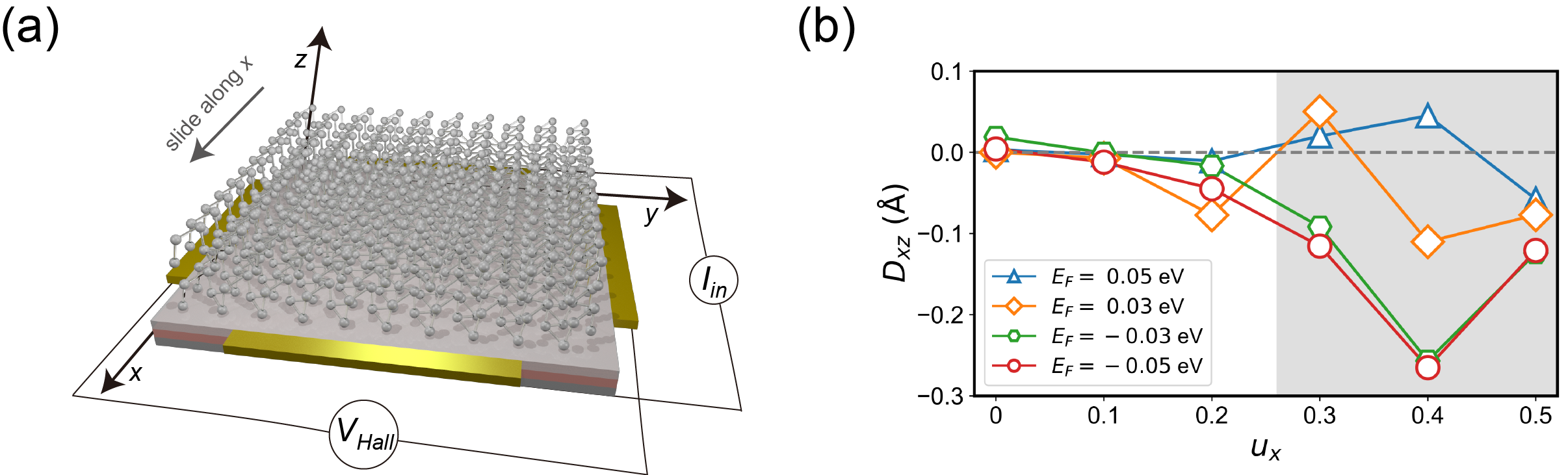}
    \caption{(a) Schematic diagram of a device consisting of bilayer bismuth for the detection of nonlinear anomalous Hall effect. (b) Berry curvature dipole component of $D_{xz}$ as a function of $u_x$ at different Fermi levels. The shaded background represents the nontrivial state. During the sliding along $x$, the absolute energy of the Fermi level is fixed.
    }
    \label{illus}
\end{figure}

Finally, we theoretically explore the sliding-driven evolution of the nonlinear anomalous Hall current, keeping the Fermi level fixed at an absolute energy. A potential device setup is depicted in Fig.\ref{illus}(a), where an input current is applied along the $x$-axis, while the transverse Hall voltage, which scales with $D_{xz}$, is measured along the $y$-axis. Figure\ref{illus}(b) illustrates the evolution of $D_{xz}$ as a function of $u_x$ for various values of $E_F$. The topologically nontrivial configurations typically exhibit larger magnitudes of $D_{xz}$, albeit the sign being sensitive to $E_F$. Experimentally, monitoring the magnitude of the Hall voltage may aid in identifying possible topological transitions.



In conclusion, we demonstrate that a sliding-induced $\mathbb{Z}_2$ topological phase transition is achievable in bilayer ferroelectric bismuth. Counterintuitively, a gentle lateral shift can result in pronounced modulations in the band structures, leading to energy gap closure and band inversion that underlie the electronic topological transition. The established topological phase diagram with respect to sliding in all possible directions in 2D reveals that sliding can drive successive trivial-nontrivial-trivial phase transitions, suggesting a deterministic, energy-efficient avenue to configure the quantum spin-Hall insulator and its symmetry-protected edge states. The spontaneous inversion symmetry breaking in ferroelectric bismuth makes it feasible to utilize the giant Berry curvature commonly associated with nontrivial topological insulators. Our DFT calculations indicate that the sign and magnitude of the Berry curvature dipole depend sensitively on the sliding direction and distance, thereby supporting a highly tunable nonlinear nomalous Hall current. We suggest that bilayer bismuth is a promising platform for manipulating two types of currents, both deeply rooted in Berry phase, for nonvolatile multistate memory applications.



\begin{acknowledgments}
Z.Q. and S.L.~acknowledge the supports from Westlake Education Foundation. The computational resource is provided by Westlake HPC Center. H.W. acknowledges the support provided by National Natural Science Foundation of China (NSFC) under Grant No.~12304049. Z.Q.~acknowledges the help from Yudi Yang and Yihao Hu during the preparation of the manuscript.
\end{acknowledgments}




\bibliography{note.bib}

\newpage


\newpage

\newpage

\clearpage
\newpage

\end{document}